\documentclass[twocolumn,preprintnumbers,amsmath,amssymb,floatfix,showpacs,prl,aps]{revtex4}

\usepackage{dcolumn}
\usepackage{bm}
\usepackage{graphicx}

\begin{document}


\title{``Hall viscosity'' and  intrinsic metric of incompressible
fractional quantum Hall fluids.}

\author{F. D. M. Haldane}
\affiliation{Department of Physics, Princeton University,
Princeton NJ 08544-0708}

\date{June 10, 2009}

\begin{abstract}
The (guiding-center) ``Hall viscosity" is a fundamental tensor 
property of incompressible ``Hall fluids'' exhibiting the 
fractional quantum Hall effect; it determines the
stress induced by a non-uniform electric field, and the intrinsic
dipole moment on (unreconstructed) edges.  
It is characterized by a rational number and an
intrinsic metric tensor that defines distances on an
``incompressibility lengthscale''.   These properties do not require
rotational invariance in the 2D plane.  The sign of the  guiding-center
Hall viscosity distinguishes particle fluids from hole fluids, and its magnitude
provides a lower bound to the coefficient of the $O(q^4)$ small-$q$ limit of
the guiding center structure factor, a fundamental measure of incompressibility.
This bound becomes an equality for conformally-invariant model wavefunctions such as Laughlin or Moore-Read states.  
\end{abstract}


\maketitle
Most treatments of the fractional quantum Hall effect (FQHE) assume  rotational invariance.   This  has  been used to demonstrate\cite{GMP} that the  ``guiding center structure factor'' $S_0(\lambda\bm q)$ of incompressible FQHE ``Hall fluids''  vanishes as $O(\lambda^4)$  as $\lambda \rightarrow 0$.   It is  also a feature of  Read's discussion\cite{read} of their  dissipationless ``Hall viscosity'', which generalizes earlier results\cite{avron} by Avron
\textit{et al.} for the integer QHE.   In this Letter, I show that the results \cite{GMP} and
\cite{read} are not only related, but can be derived from translational invariance  without invoking rotational invariance.    Both properties are  characterized by  tensors  that can be obtained from the response of FQHE wavefunctions with periodic boundary conditions (pbc's) to adiabatic changes in pbc geometry.   The physical significance of the ``Hall viscosity tensor'' emerges: it characterizes an intrinsic electric dipole moment on unreconstructed edges of the  Hall fluid, and provides an intrinsic  metric deriving from the shape of the correlations that give rise to incompressibility.

A ``generic'' Hall system is formed by a 2D electron gas (2DEG) 
bound to a flat 2D quantum well  embedded in a uniform 3D dielectric, through which a
magnetic flux passes, with a uniform flux  density $\Phi_0 /2\pi \ell^2$ , where $\Phi_0$ =
$2\pi \hbar/e$.    Provided the ``magnetic length'' $\ell$ is much larger than atomic
lengthscales, the clean system can be regarded as translationally-invariant in the two directions parallel to the plane of the quantum well.   The one-particle eigenstates of an electron in the well have the form
\begin{equation}
H_0^{(1)}|\psi_{n\alpha}\rangle = \varepsilon_n |\psi_{n\alpha}\rangle ,
\end{equation}
where $n$ is an index that combines quantum well, valley, spin, and Landau level indices.  Each of these \textit{generalized Landau levels} is a macroscopically-degenerate multiplet (with degeneracy $N_{\rm orb}$, the 2D area in units $2\pi\ell^2$)  in which the
``magnetic translation group'' (guiding center algebra) acts:
\begin{eqnarray}
&&\sum_{\beta}
\langle \psi_{n\alpha}|e^{i\bm q\cdot \bm R} |
\psi_{n\beta}\rangle
\langle \psi_{n\beta}|e^{i\bm q'\cdot \bm R} |
\psi_{n\gamma}\rangle = \nonumber \\
&&\quad e^{i\frac{1}{2}\bm q\times \bm q' \ell^2}
\langle \psi_{n\alpha}|e^{i(\bm q+\bm q')\cdot \bm R} |
\psi_{n\gamma}\rangle.
\end{eqnarray}
Here $\bm q\times \bm q'$ $\equiv$ $\epsilon^{ab}q_aq'_b$, where $\epsilon^{ab}$ =
$\epsilon_{ab}$ is the 2D antisymmetric symbol, with an orientation (chirality) defined by the magnetic vector potential
in the plane of the 2DEG:
$e(\nabla_aA_b(\bm r)$ $-$ $\nabla_bA_a(\bm r))$ = $\hbar \epsilon_{ab}/\ell^2$.
The dynamical momentum of electrons in the 2DEG is
$\pi_{ia}$ $\equiv$  $p_{ai} - eA_a(\bm r_i)$, 
with $[r_i^a,p_{jb}]$ = $i\hbar\delta_{ij}\delta^a_b$, and
$[r^a_i,r^b_j]$ = 0 = $[p_{ia},p_{jb}]$.
$R^a_i$ = $r^a_i - (\ell^2/\hbar)\epsilon^{ab}\pi_{ib}$ are the \textit{guiding centers}
that commute with the dynamical momenta, and obey the algebra
$[R^a_i,R^b_j]$ = $i\ell^2\delta_{ij}\epsilon^{ab}$.
Note the use of a  covariant 
formalism  where 2D spatial coordinates $r^a$ have upper indices $a$ = 1,2, and
reciprocal vectors such as wavenumbers $q_a$ have lower indices, and only upper/lower pairs can be contracted: $\bm q \cdot \bm r$ $\equiv$ $q_ar^a$; this formalism makes the metric-independence of generic  FQHE properties explicit.

I will assume that the low-energy effective Hamiltonian that describes degenerate perturbation theory in the residual two-body Coulomb interaction conserves
the number of electrons in each level $\varepsilon_n$: this will be true if the splitting of the levels is large compared to the interaction strength, or if no spin-orbit coupling or
interlayer tunneling mixes weakly-split levels.

The only natural metric in the problem is provided by the long-distance  behavior of the non-relativistic (non-retarded) Coulomb interaction
\begin{equation}
V_C(\bm r_i-\bm r_j) = \frac{e^2}
{4\pi\varepsilon_0\bar\varepsilon |\bm r_i - \bm r_j|_C},
\quad 
|\bm r|^2_C \equiv \bar\varepsilon \varepsilon^{-1}_{ij}r^ir^j,
\end{equation}
where $\varepsilon^{ij}$ is the dielectric tensor of the 3D medium surrounding the
2DEG, and $\bar \varepsilon $ = $(\det |\varepsilon|)^{1/3}$;  $|\bm r- \bm r'|_C$ can 
be called the ``Coulomb distance'' between two points, and Coulomb distances
between points on the 2D plane are given by
$|r|_C$ = $|r|_g$ $\equiv$ $(g_{ab}r^ar^b)^{1/2}$, where
$g_{ab}$ is a positive-definite 2D metric tensor with $\det |g|$ = 1. 

However, the long-distance part of the Coulomb interaction is not the source of incompressibility of  FQHE Hall fluids, and is a complication that must be treated separately: on large lengthscales, it forces local charge-neutrality on the system, which generally requires  a finite density of charged vortices (quasiparticles).
As an ``action at a distance'', it also violates the (quasi-)local continuity equation for momentum.

FQHE incompressibility is analogous to  Mott-Hubbard incompressibility, where
an energy  gap ``$U$''  prevents a second particle from occupying
an already-occupied atomic orbital.  
In the $\nu$ = 1/3 Laughlin state $|\Psi_L^{1/3}(g)\rangle$ \cite{laughlin},  
localized 
orbitals $|\psi_{nm}(\bm r,g)\rangle$ of level $n$, centered at $\bm r$, are  eigenstates with eigenvalue $(m+\frac{1}{2})\hbar$,  $m\ge 0$,  of 
\begin{equation}
L_i^z(\bm r, g)  =
\frac{\hbar}{2\ell^2}g_{ab}(R_i^a-r^a)(R_i^b-r^b), 
\end{equation}
where
$g_{ab}$ is \textit{any}  metric tensor  and is a tunable  parameter.
In the Mott-Hubbard analogy, an energy gap prevents occupation of
$|\psi_{n1}(\bm r,g)\rangle$  if $|\psi_{n0}(\bm r,g)\rangle$  is already occupied.
FQHE incompressibility in general can be understood as the exclusion of additional particles from a region already occupied by a certain cluster of particles with a
specific shape (controlled by  the ``hidden''  variational parameter $g$ in the Laughlin state).

It  is convenient to
separate short-range and long-range parts by writing the Coulomb interaction
as the sum of a  long-range part $V_C(|\bm r|_C^2 + a^2)^{1/2})$ and a short
range part  $V_0(\bm r,a)$ which is the  interaction screened by a fictitious conducting plane parallel to the 2DEG, set back a (Coulomb) distance $a/2$ from it.   The long-range part
cancels the fictitious image-charge  potentials.   The distance $a$ should be larger than the intrinsic ``incompressibility lengthscale'', but is otherwise arbitrary.

The 2D electron density is related to the 3D
electron density by $\rho(\bm r)$ = $\int dz \, \rho^{(3D)}(\bm r,z)$,
and given by
\begin{equation}
\rho(\bm r) =
\sum_{n}
\int \frac{d^2q}{(2\pi)^2}e^{-i\bm q\cdot \bm r}
f_n(\bm q) \langle \bar \rho_{n}(\bm q) \rangle .
\end{equation}
Here $f_n(\bm q)$  (where $f_n(0)$ = 1) is a form factor associated with the one-particle wavefunctions of
level $n$, and $\bar \rho_n(\bm q)$ is the Fourier-transformed 
\textit{guiding-center density}
\begin{equation}
\bar \rho_n(\bm q) = \sum_i P^n_ie^{i\bm q \cdot \bm R_i},
\end{equation}
where $P^n$  is the projection into level $n$. 
If (as usually assumed) there is isotropy and Galilean invariance in the plane of
the 2DEG,
\begin{equation}
f_n(\bm q ) = 
L_{n}({\textstyle\frac{1}{2}}|q|^2_g\ell^2)
e^{-\frac{1}{4}|q|^2_g\ell^2},
\end{equation}
where $n$ is the Landau index and $L_n$ are Laguerre polynomials.
Here the metric that defines $|q|_g$ derives from the \textit{Galilean effective mass tensor}
$m^*g_{ab}$ and only coincides with the Coulomb metric 
if there is isotropy in the 2D plane,
but if isotropy is broken (\textit{e.g.}, by ``tilting'' the magnetic field),  is distinct from
it.

The  guiding-center densities obey the Lie algebra\cite{GMP}
of generators of ``diffeomorphisms of the quantum plane''
\begin{equation}
{[}\bar \rho_n(\bm q),\bar \rho_{n'}(\bm q')]
= 2i\delta_{nn'}\sin ({\textstyle\frac{1}{2}} \bm q\times \bm q' \ell^2)
\bar \rho_n(\bm q+ \bm q').
\label{GMPlie}
\end{equation}
In a uniform Hall fluid
$\langle \bar \rho_n(\bm q)\rangle$ = $2\pi \nu_n\delta^2(\bm q\ell)$,
where $\nu_n$ is the filling factor of level $n$.   The regularized densities
$\bar \rho_n(\bm q) - \langle \bar \rho_n(\bm q)\rangle $ also
obey (\ref{GMPlie}), with
$\lim_{\lambda\rightarrow 0}\bar \rho_n(\lambda \bm q)$ = 0.     
 Two important subalgebras are  those of the
generators of translations and linear deformations of the plane:
If $\bar \rho(\bm q)$ = $\sum_n\bar \rho_n(\bm q)$ 
\begin{eqnarray}
&\lim_{\lambda\rightarrow 0}& \nabla_q^a\bar \rho(\lambda\bm q)
\rightarrow
- i\lambda \ell^2\epsilon^{ab}\hbar^{-1}P_b ,\\
&\lim_{\lambda\rightarrow 0} &
{\textstyle\frac{1}{2}}
\nabla_q^a\nabla_q^b\bar \rho(\lambda\bm q)
\rightarrow
 (i\lambda \ell)^2\Lambda^{ab}.
\end{eqnarray}
Note that $[P_a,P_b]$ = $\bar\rho(0)(\hbar^2/\ell^2)\epsilon_{ab}$, which vanishes
if $\bar \rho (\bm q)$ is the regularized density, so the uniform  ground state
of the Hall fluid consistently obeys $P_a|\Psi_0\rangle$ = 0.
The three independent components $\Lambda^{ab}$
= $\Lambda^{ba}$ are generators
of  unitary transformations 
$U(\alpha)$ = $\exp i \alpha_{ab}\Lambda^{ab}$, where
$U(\alpha)R^a_iU(-\alpha)$ = $\lambda^a_b(\alpha) R_i^b$ is a linear
transformation  of guiding centers that preserves
their algebra;
they satisfy the $SO(2,1)$ Lie algebra
\begin{eqnarray}
[\Lambda^{ab},\Lambda^{cd}]
&=& {\textstyle\frac{1}{2}}i\left ( 
\epsilon^{ac}\Lambda^{bd} +
\epsilon^{ad}\Lambda^{bc} +
\epsilon^{bc}\Lambda^{ad} +
\epsilon^{bd}\Lambda^{cd} \right ) \quad \quad
\end{eqnarray}
with  quadratic Casimir   $C_2$ =  $-\det |\Lambda |$.
If there is rotational invariance with metric $g$,
\begin{equation}
L^z(g) \equiv L^z(0,g) = \hbar g_{ab}\Lambda^{ab}.
\end{equation}

Now consider an incompressible Hall fluid subject to a slowly varying local potential $V(\bm r)$, with the
Hamiltonian 
\begin{equation}
H=  H_0 + V_1 \equiv \int d^2\bm r \, h_0(\bm r) + V(\bm r)  \bar\rho(\bm r),
\end{equation}
where $h_0(\bm r)$ is the quasilocal part of the  Hamiltonian derived from the
cyclotron motion kinetic energy and short-range part of of the Coulomb interaction,
and $V(\bm r)$ is the local potential derived from both external  potentials
and the long-range part of the Coulomb interaction.    Here, it is assumed that
$V(\bm r)$ and $\rho(\bm r) $ $\equiv$ $\langle \bar \rho(\bm r)\rangle $ are slowly varying, so all significant
Fourier components have  $|q|_C\ell \ll 1$.
The local continuity relations for particle density $\rho(\bm r)$ and momentum density
$\pi_a(\bm r)$ are
\begin{equation}
\partial_t\rho + \nabla_aj^a = 0,
\quad
\partial_t\pi_a  + \nabla_b\sigma_a^b + \rho \nabla_a V = 0,
\end{equation}
where $j^a(\bm r)$ is the current density and $\sigma^a_b(\bm r)$ is the local stress tensor.  Note that local momentum conservation is violated by the  ``body-force''
field $\nabla_a V(\bm r)$.  

The effect of $V(\bm r)$ can be treated in linear
response inside a bulk region of the fluid where $|V(\bm r) - V(\bm r')|$ $ <$ $\Delta$,
the threshold for exciting topological excitations.
For $\nu$ = $\sum_n\nu_n$, 
\begin{eqnarray}
j^a &=& \frac{\nu}{2\pi \hbar}\epsilon^{ab}\nabla_{b}V, \quad
\sigma^a_e = \frac{1}{2\pi} \epsilon_{eb}\Gamma_A^{abcd}\nabla_c\nabla_dV .
\label{response}
\end{eqnarray}
The pressure $p(\bm r)$  = $\sigma^a_a(\bm r)$ vanishes inside the
bulk regions of the Hall fluid, because  the edge states shield them  from forces applied to the edge: if the ``container'' is squeezed, the edge-current  increases, as does the potential
at the edge, but provided the condition $|V(\bm r) - V(\bm r')|$ $<$ $\Delta$
remains true inside the fluid, no
effects  are felt in the interior.  The  condition $p(\bm r)$ = 0 implies
$\Gamma_A^{abcd}$ = $\Gamma_A^{bacd}$. 

The stress tensor defines the force $dF_a$ = $\sigma^b_a\epsilon_{bc}dL^c$  between
the regions of fluid on either side of a cut along a line segment $d\bm L$.
The absence of dissipation in the ground state of the fluid in the presence of
the potential $V(\bm r)$ imposes the condition  $\sigma^a_b\nabla_aj^b$ = 0,
or $\Gamma_A^{abcd}$ = $-\Gamma_A^{cdab}$.
Therefore $\Gamma_A^{abcd}$ has the form
\begin{equation}
\Gamma_A^{abcd} = \pi\left (
\epsilon^{ac}Q^{bd} +
\epsilon^{ad}Q^{bc} +
\epsilon^{bc}Q^{ad} +
\epsilon^{bd}Q^{ac} \right ),
\end{equation}
where $Q^{ab}$ = $Q^{ba}$ is \textit{symmetric}.
The expression (\ref{response}) can be reformulated as a ``Hall viscosity'' relation
by introducing the drift velocity field $v^a(\bm r)$
 = $\hbar^{-1}\ell^2\epsilon^{ab}\nabla_bV$:
then $\sigma^a_b$ = $\eta^{ac}_{bd}\nabla_cv^d$, where
$\eta^{ac}_{bd}$ = $(\hbar/2\pi\ell^2)\epsilon_{be}\epsilon_{df}\Gamma_A^{aecf}$.

The tensor $Q^{ab}$ also has the physical significance of describing the 
\textit{intrinsic electric dipole moment} on the (unreconstructed) boundary of a Hall fluid in its ground state,
or, more generally, on the boundary between two Hall fluids with different intrinsic
tensors $Q^{ab}$.
 If $d\bm L$ is an infinitesimal
line segment on the (static) boundary,
it must follow an equipotential: $\bm \nabla V \cdot d\bm L $ = 0.
The discontinuity in the stress tensor field  leaves a net 
stress force on the boundary, where $dF_a$ = $\Delta \sigma^b_adL^b$ =  
$-dp^b\nabla_b E_a$, where $eE_a$ = $-\nabla_aV$,  and
\begin{equation}
dp^a 
= e\Delta Q^{ab}\epsilon_{bc}dL^c.   
\label{dipole}
\end{equation}
For the boundary to remain static, $d\bm p$ must correspond precisely
to an intrinsic
electric dipole that experiences a compensating force from the electric field
gradient to balance the stress discontinuity,  and
(\ref{dipole}) may be regarded as a fundamental relation of Hall fluids.
Stability requires the dipole to point in the direction
of decreasing electric charge density, so
$\Delta \nu \Delta Q^{ab}$ is positive.

This allows a direct calculation of $\Delta Q^{ab}$ from the edge properties.
A ``Landau gauge'' basis of one-particle eigenstates $\bm n \cdot \bm R|\Psi_n(k)\rangle$
= $k\ell|\Psi_n(k)\rangle $ is used to describe a translationally-invariant straight boundary along the line $\bm n \cdot \bm r $ = 0.     Occupation numbers $\nu_n(k)$ vary continuously between limits $\nu_n^{(\pm)}$ for $\bm n\cdot\bm r$ $\rightarrow$ $\pm \infty$, where $k$ is proportional to the distance from the boundary.
The total occupation number $\nu(k)$ = 
$ \sum_n \nu_n(k)$  has a singularity 
at $k$ = 0 (with a form predicted by the conformal field theory of the
edge\cite{wen}), because  an electron can be locally added gaplessly at  positions on
the boundary.  The dipole moment per unit length of wall is the sum of
two terms: an ``integer QHE'' term derived completely from the form 
factor $f_n(\bm q)$ and the discontinuities $\nu_n^{(+)}-\nu_n^{(-)}$,
and a ``fractional QHE'' term that is independent of the form factor,
and depends on the functions $\theta(\pm k)(\nu(k)-\nu^{(\pm)} )$, which vanish in the integer QHE case.
Then
\begin{equation}
Q^{ab} = -\frac{1}{4\pi\ell^2}\sum_n\nu_n\nabla_q^a\nabla_q^bf_n(\bm q)
+ Q_0^{ab},
\end{equation}
where $Q_0^{ab}$ is the  residual guiding-center contribution obtained from
the sum rule
\begin{equation}
\int_0^{\infty} \frac{dk}{2\pi}(\alpha + \beta k)(\nu(k)-\nu(\infty))
= \beta Q^{ab}_0n_an_b.
\label{sumrule}
\end{equation}
$Q_0^{ab}$  cannot change adiabatically as $H_0$ changes, because it is fixed by
the conserved momentum parallel to the edge.  Note that (\ref{sumrule})
is unchanged by rescaling $n_a$ $\rightarrow$
$\lambda n_a$, so is correctly metric-independent. Adiabatic continuity of the edge
properties as $n_a$ changes appears to require that the sign of $Q_0^{ab}n_an_b$ cannot change, which implies that $Q_0^{ab}$ is positive or negative definite: $\det |Q_0| > 0$.

Dynamical edge modes are described\cite{wen} by one or more
conformal field theories with a net chiral anomaly $\Delta \nu$.
Excitation of these modes changes the magnitude of the edge dipole moment
by an amount proportional to the linear momentum (Virasoro level) of the excitation,
and it will be  temperature-dependent in a non-universal way that depends
on the (local) edge-mode velocities.
A ground-state instability (reconstruction) of the edge can also change the dipole
moment.

The fractional part $Q^{ab}_0$  can also be derived from the expression for the
guiding-center stress tensor, obtained  by
solving the operator relation $\hbar q_b\sigma^b_a(\bm q)$ = $[\pi_a(\bm q),H_0]$,
(using $[\pi_a(0),H_0]$ $\equiv$  $[P_a,H_0]$ = 0),
unlike previous  metric-based derivations\cite{vignale} which appear to
assume rotational invariance. 
Here $\pi_a(\bm q)$ = $(\hbar/\ell^2)\epsilon_{ab} [R^b]_{\bm q}$,
where I define
the generalized Fourier-transformed density
\begin{equation}
 [f(\bm R)]_{\bm q} \equiv 
\sum_i e^{\frac{1}{2}i\bm q\cdot \bm R_i}
f(\bm R_i) e^{\frac{1}{2}i\bm q\cdot \bm R_i},
\end{equation}
where \textit{e.g.}, $\bar \rho (\bm q)$
$\equiv$ $\sum_n\bar\rho_n(\bm q) $ =
$[1]_{\bm q}$.    After some simple manipulations, using
$\langle  [X,H_0] \rangle_V$ = $-\langle [X,V_1]\rangle_0 $ + $O(V^2)$
(where $\langle \ldots \rangle_V$ is the ground state expectation value
in the non-uniform system perturbed by $V_1$),
I obtain the formal results
\begin{eqnarray}
Q_0^{ab} &=& \frac{1}{N_{\rm orb}}\frac{\langle \Lambda^{ab}\rangle_0}{4\pi}, 
\quad \Gamma_{0A}^{abcd}  = \frac{-i}{N_{\rm orb}}
\langle{\textstyle\frac{1}{2}} [\Lambda^{ab},\Lambda^{cd}]\rangle_0, \quad \quad
\end{eqnarray}
where  $\Lambda^{ab}$ is to be understood as  the regularized quantity $[\Lambda^{ab}]_0$ =
$\lim_{\lambda \rightarrow 0}[\Lambda^{ab}]_{\lambda\bm q}$.

These results are in complete agreement with the earlier results\cite{avron,read}
for the case where $L^z(g)$ is the generator of a rotational symmetry, and $g^{ab}$
is the Galilean tensor of the cyclotron orbits: 
$\hbar Q^{ab}/\ell^2 $ has the form\cite{avron,read} $\eta^{(A)}g^{ab}$, 
\begin{equation}
Q^{ab}  =\frac{g^{ab}}{4\pi}
\sum_n (\nu_ns_n + \frac{\gamma_n }{2} ), \quad \gamma_n \equiv
N_n - \nu_nN_{\rm orb} ,
\end{equation}
where $s_n$ = $n+\frac{1}{2}$.    
Here $\gamma$ = $\sum_n\gamma_n$ is a ``shift'' (here defined ``per flux'', rather than ``per particle'') 
seen in the polynomial wavefunctions that describe rotationally-invariant Hall fluids in disk or sphere geometries, and 
is a purely  guiding-center quantity that vanishes for a completely-filled Landau level,
and is odd under particle-hole conjugation.   
If $\nu$ = $p/q$,  then
$q\gamma $ is an integer.
Without regularization,
the guiding center contribution  $g_{ab}\Lambda^{ab}$ to
$L^z/\hbar$  for a circular finite-size Hall droplet is $\frac{1}{2} NN_{\rm orb}$, but regularization removes the superextensive part $\frac{1}{2}\nu N^2_{\rm orb}$, leaving the (extensive)
regularized part $\frac{1}{2}\gamma N_{\rm orb}$.
For the $\nu$ = $k/(km+2)$
Laughlin\cite{laughlin} ($k=1$), Moore-Read\cite{MR} ($k=2)$, or Read-Rezayi\cite{RR} ($k > 2$) states in a single Landau level, $\gamma$ = $\nu (m+1)$,
($m$ is odd for fermions, even for bosons).

If a single Hall fluid is present, the condition
$\det |Q_0| > 0$ implies  $Q^{ab}_0$ = $\pm {\textstyle\frac{1}{2}}|\gamma |g^{ab}/4\pi$, where $|\gamma |$
is positive rational, $g^{ab}$ is an intrinsic metric \textit{defined} by the incompressibility, and the sign of $Q_0^{ab}$ distinguishes 
electron Hall fluids
$(Q_0^{ab} >0)$ from  hole Hall fluids ($Q_0^{ab}<0$).  
If there are multiple (independently deformable) Hall fluids,
$Q^{ab}_0$ will be the sum of  tensors 
associated with each fluid.

I now make the connection to the long-wavelength behavior of the guiding-center structure factor, given in terms of the total (regularized)  guiding-center density
operators by
\begin{equation}
\langle \bar \rho(\bm q)\bar \rho(\bm q')\rangle_0 =
2\pi S_0(\bm q)\delta^2(\bm q\ell +\bm q'\ell),
\end{equation}
or in the unregularized form,
\begin{equation}
S_0(\bm q)  = \frac{1}{N_{\rm orb}}
\langle \bar \rho(\bm q)\bar\rho(-\bm q)\rangle_0
-
\langle \bar \rho(\bm q)\rangle_0\langle\bar\rho(-\bm q)\rangle_0.
\end{equation}
Note that this is normalized per flux rather than per particle, and
is even under a particle-hole transformation of the partially-filled
Landau levels (and vanishes identically if all Landau levels are either filled or empty).
$S_0(\lambda \bm q)$ vanishes as $\lambda $ $\rightarrow$ 0 (because the total particle number is fixed), and is even in $\lambda$.   Since the ground state is assumed to be  incompressible, with a gap for excitations, $S_0(\lambda \bm q)$
will be analytic in  $\lambda$, and the formal expansion coefficients depend on the fluctuations $C_{mn}$ = 
$\langle [(\bm q\cdot \bm R)^m]_0 [(\bm q\cdot \bm R)^n]_0\rangle_0$
$- $ $\langle [(\bm q\cdot \bm R)^m]_0\rangle_0\langle [(\bm q\cdot \bm R)^n]_0\rangle_0$.
The crucial point is that $[R^a]_0$
= $\ell^2\epsilon^{ba}P_b/\hbar$ which is conserved, so all coefficients $C_{1m}$ must vanish.   This
argument asserts incompressibility and translational invariance of the uniform Hall fluid,
with \textit{no} requirement of rotational invariance.   The leading behavior of 
$S_0(\lambda \bm q)$ at small $\lambda$ is derives from $C_{22}$:
$S_0( \lambda \bm q)$ = $\frac{1}{4}\lambda ^4\Gamma_{0S}^{abcd}q_aq_bq_cq_d\ell^4$ + $O(\lambda^6)$,
where $\Gamma^{abcd}_{0S}$  is the symmetric
fourth-rank tensor 
\begin{equation}
\Gamma_{0S}^{abcd} = \frac{1}{N_{\rm orb}}
\left (
\langle {\textstyle\frac{1}{2}}\{\Lambda^{ab},\Lambda^{cd}\}
\rangle_0 -
\langle\Lambda^{ab}\rangle_0
\langle\Lambda^{cd}\rangle_0\right ).
\end{equation}
Evidently this can be combined with $\Gamma_{0A}^{abcd}$ to form the complex Hermitian
tensor $\Gamma_0^{abcd}$=  $\Gamma_{0S}^{abcd} $+ $i\Gamma_{0A}^{abcd} $, \textit{i.e.}:
\begin{equation}
\Gamma^{abcd}_0 =
\frac{1}{N_{\rm orb}}\left (
\langle \Lambda^{ab}\Lambda^{cd}
\rangle_0 -
\langle\Lambda^{ab}\rangle_0
\langle\Lambda^{cd}\rangle_0
\right ) .
\end{equation}

The linear deformations  that preserve the guiding-center algebra
can be written 
$U(\bm x)$ = $exp \left (i \alpha_{ab}(\bm x)\Lambda^{ab}\right )$, where
$\bm x$  = $\{x^{\mu}, \mu =1,2,3\}$ is a three-dimensional manifold
corresponding to the three distinct generators $\Lambda^{ab}$.
 Let $|\Psi(\bm x)\rangle$ = $U(\bm x)|\Psi_0\rangle$ be a deformation of the incompressible fluid ground state.  The covariant derivative in the deformation parameter
space $\bm x$ is given by
\begin{equation}
|D_{\mu}\Psi\rangle = \left ( \openone - |\Psi\rangle\langle \Psi |\right )
\left |\frac{\partial \Psi}{\partial x^{\mu}} \right \rangle,
\quad \langle \Psi|D_{\mu}\Psi\rangle = 0.
\end{equation}
Then $\langle D_{\mu}\Psi|D_{\nu}\Psi\rangle$ = 
$\Gamma_0^{abcd}\partial_{\mu}\alpha_{ab}
\partial_{\nu}\alpha_{cd} $
is a $3 \times 3$ positive Hermitian matrix which can be obtained by studying the adiabatic
deformation of a finite-size Hall fluid system with periodic boundary conditions as
pbc geometry is varied; from this, $\Gamma_0^{abcd}$ can be determined\cite{unpub}.
This generalizes the two-parameter  deformation space studied in  Refs.
\cite{avron,read} where  rotational invariance was assumed.

If the fluid ground state is an eigenstate of $L^z( g)$, then $g_{ab}\Gamma_0^{abcd}$
must vanish, which forces $\Gamma_{0S}^{abcd}$ to have the structure
$\frac{1}{2}\kappa(g^{ac}g^{bd} + g^{ad}g^{bc} - g^{ab}g^{cd})$, where $g^{ab}$ is the metric also defined by $Q_0^{ab}$.     $\Gamma^{abcd}$ can viewed as a positive 
 Hermitian matrix $\Gamma^{\{ab\},\{cd\}}$,
 which leads to the bound $\kappa$ $\ge$
$|\gamma|$ (because $\Gamma_S^{\{ab\},\{cd\}} \pm i\Gamma_A^{\{ab\},\{cd\}}$ is positive).  
The value of $\kappa$ is known\cite{GMP} for the
Laughlin states and \textit{satisfies the bound as an equality} $\kappa$ = $|\gamma|$. 
Numerical diagonalization studies\cite{unpub}  show this is also true
for the finite-$N$ Moore-Read states\cite{MR}, suggesting it is a general
property of maximally-chiral 
model wavefunctions derived from conformal field theory.  Corrections
to these wavefunctions when realistic interactions are used
appear\cite{unpub} to generically increase $\kappa$ above the bound,  but it is not yet clear whether or not this is a finite-size effect.

This work was supported in part by the U. S. National
Science Foundation (under MRSEC Grant No. DMR-0819860) at  the Princeton
Center for Complex Materials..

\end{document}